\documentclass[twocolumn,showpacs,aps,prl,superscriptaddress]{revtex4}
\usepackage{graphicx}
\usepackage{dcolumn}
\usepackage{epsfig}
\usepackage{amsmath}
\RequirePackage{xspace}

\def\epem  {\ensuremath{e^+e^-}\xspace}

\newcommand{\thetaT}{\ensuremath{\theta_{\rm T}}}
\newcommand{\costhr}{\ensuremath{\cos\thetaT}}

\newcommand{\xf}{\mbox{${\cal F}$}}

\def\Bu      {\ensuremath{B^+}}
\def\Bub     {\ensuremath{B^-}}
\def\Bbar    {\overline{B}{}}

\def\Bzb     {\ensuremath{\Bbar^0}}
\def\Bz      {\ensuremath{B^0}}
\def\BpBm    {\ensuremath{\Bu  \Bub}}
\def\BzBzb   {\ensuremath{\Bz  \Bzb}}
\newcommand{\DE}{\ensuremath{\Delta E}}
\newcommand{\pvec}{{\bf p}}
\newcommand{\half}{\mbox{${1\over2}$}}
 \def\mes{\mbox{$m_{\rm ES}$}}
\newcommand{\calB}{\mbox{${\cal B}$}}
\newcommand{\auno}{\mbox{$a_1(1260)$}}

\newcommand{\Nappim}{\mbox{$a^{\pm}_1(1260)\, \pi^{\mp}$}}

\newcommand{\Nadueppim}{\mbox{$a^{\pm}_2(1320)\, \pi^{\mp}$}}

\newcommand{\Npipppim}{\mbox{$\pi^{\pm}(1300)\,\pi^{\mp}$}}

\newcommand{\bmtoappim}{\mbox{$B^0 \rightarrow a^-_1(1260)\, \pi^+  $}}
\newcommand{\Nbtoappim}{\mbox{$B^0 \rightarrow a^{\pm}_1(1260)\, \pi^{\mp}  $}}
\newcommand{\Nbtoadueppim}{\mbox{$B^0 \rightarrow a^{\pm}_2(1320)\, \pi^{\mp}  $}}
\newcommand{\Nbtopipppim}{\mbox{$B^0 \rightarrow \pi^{\pm}(1300)\, \pi^{\mp}  $}}

\newcommand{\NNatorhopi}{\mbox{$ a_1(1260) \rightarrow \rho \pi  $}}

\newcommand{\Natopipipi}{\mbox{$ a^{\pm}_1(1260) \rightarrow \pi^{\mp}\pi^{\pm}\pi^{\pm}  $}}

\newcommand{\UfourS}{\mbox{$\Upsilon(4S)$}}

\newcommand{\BrapiPiPiPi}{\mbox{$\calB(\Nbtoappim) \calB(\Natopipipi)$}}

\newcommand{\rapi}{\mbox{$16.6 \pm 1.9 \pm 1.5 $}}
\newcommand{\Rapi}{\mbox{$(\rapi)\times 10^{-6}$}}

\def\babar{{\em B}{\footnotesize\em A}{\em B}{\footnotesize\em AR}}
\def\BB{\mbox{$B\overline B\ $}}
\def\pep2{PEP-II}

\newcommand\etal{{\it et al.}}
\newcommand{\dedx}{\ensuremath{\mathrm{d}\hspace{-0.1em}E/\mathrm{d}x}}
\newcommand{\gevcc}{\mbox{$\textrm{GeV}/c^2$}} 
\newcommand{\mevcc}{\mbox{$\textrm{MeV}/c^2$}} 
\newcommand{\gevc}{\mbox{$\textrm{GeV}/c$}} 
\newcommand{\gev}{\mbox{$\textrm{GeV}$}} 
\newcommand{\mev}{\mbox{$\textrm{MeV}$}} 
 
\newcommand{\jprlBase}  [1]     {Phys.\ Rev.\ Lett.}
\newcommand{\jprl}      [1]    {\jprlBase\ B~{\bf #1}}
\newcommand{\jprBase}        {Phys.\ Rev.\ }
\newcommand{\jprd}      [1]  {\jprBase\ D~{\bf #1}}
\newcommand{\plBase}   [1]         {Phys.\ Lett.}
\newcommand{\plb}      [1]    {\plBase\ B~{\bf #1}}
\newcommand{\nimBaseA}       {Nucl.\ Instr.\ Meth.\ }
\newcommand{\nima}      [1]  {\nimBaseA~A~{\bf #1}}
\newcommand{\zpBase}         {Z.\ Phys.}
\newcommand{\zpc}       [1]  {\zpBase\ C~{\bf #1}}
\newcommand{\npBase}         {Nucl.\ Phys.\ }
\newcommand{\npb}       [1]  {\npBase\ B~{\bf #1}}

\def\figurebox#1#2#3{%
    \def\arg{#3}%
    \ifx\arg\empty
    {\hfill\vbox{\hsize#2\hrule\hbox to #2{\vrule\hfill\vbox to #1{\hsize#2\vfill}\vrule}\hrule}\hfill}%
    \else
    {\hfill\epsfbox{#3}\hfill}%
    \fi}

\begin{document}

%\begin{flushleft}
% Take this out for PUB  ----------------------------------
%     \rm \babar$\;$Analysis Document \#1311 
%      Version 18 \\
%      \today \\[.7in]
%----------------------------------------------------------
%\end{flushleft}

%\preprint{\babar-PUB-\BaBarYear/\BaBarNumber} 
%\preprint{SLAC-PUB-\SLACPubNumber} 

%\begin{flushleft}
%\babar-PUB-\BaBarYear/\BaBarNumber \\
% SLAC-PUB-\SLACPubNumber\\
% \end{flushleft}

\title{\large  \bf\boldmath  Observation of $B^0$ Meson Decay to \Nappim}

%% author list as of 08-Feb-2006 (620 authors)
%
\author{B.~Aubert}
\author{R.~Barate}
\author{M.~Bona}
\author{D.~Boutigny}
\author{F.~Couderc}
\author{Y.~Karyotakis}
\author{J.~P.~Lees}
\author{V.~Poireau}
\author{V.~Tisserand}
\author{A.~Zghiche}
\affiliation{Laboratoire de Physique des Particules, F-74941 Annecy-le-Vieux, France }
% ...
\collaboration{The \babar\ Collaboration}
\noaffiliation

\begin{abstract}

We present a measurement of the branching fraction of the decay 
\Nbtoappim\  with \Natopipipi. The data sample corresponds to $218
\times 10^6$ \BB\ pairs  produced in \epem\ annihilation through the
\UfourS\  resonance. We measure the branching fraction
\BrapiPiPiPi=\Rapi, where the first error quoted is statistical and 
the second is systematic.
\end{abstract}


\maketitle

The rare decay  \Nbtoappim\ is
expected to be dominated by $b \rightarrow u \bar{u} d$ contributions. 
For the branching fraction of this decay mode an upper limit
of $49 \times 10^{-5}$ at the 90\% C.L. has been set by CLEO~\cite{CLEO}.
Bauer {\em et al.} have predicted a branching fraction of the decay
\bmtoappim\ of $38 \times 10^{-6}$ within the framework of the
factorisation model and assuming $\left|V_{ub}/V_{cb}\right|$ = 0.08
\cite{Bauer}.
The study of this decay mode is complicated by the large discrepancies
between the parameters of the $a_1(1260)$ meson obtained from analyses
involving hadronic interactions~\cite{palombo} and $\tau$
decays~\cite{tau}.
The decay  \Nbtoappim, in addition to the decays
$B^0 \rightarrow \pi^+ \pi^-$, $B^0 \rightarrow \rho^{\pm} \pi^{\mp}$, and 
$B^0 \rightarrow \rho^+ \rho^-$
, can be used to give a new measurement of
the Cabibbo-Kobayashi-Maskawa angle $\alpha$ of the Unitarity Triangle~\cite{aleksan}.

We present a measurement of the branching fraction of the decay
\Nbtoappim\   with \Natopipipi.
The $a_1(1260) \rightarrow 3 \pi$ decay proceeds mainly through the intermediate states
$(\pi \pi)_{\rho} \pi$ and $(\pi \pi)_{\sigma} \pi$ \cite{PDG2004}.
No attempt is made to separate  the contributions
of the dominant P-wave $(\pi \pi)_{\rho}$ 
and the S-wave $(\pi \pi)_{\sigma}$ in the channel $\pi^+ \pi^-$.
Only a systematic uncertainty is estimated due to the difference in 
the selection efficiency.
Possible background contributions from $B^0$ decays to 
\Nbtoadueppim\ and \Nbtopipppim\ are investigated.

The data were collected with the \babar\ detector~\cite{BABARNIM}
at the PEP-II asymmetric $e^+e^-$ collider~\cite{pep}. An integrated
luminosity of 198~fb$^{-1}$, corresponding to
218 million \BB\ pairs, was recorded at the $\Upsilon (4S)$ resonance
(``on-resonance'', center-of-mass energy $\sqrt{s}=10.58~\gev$).
An additional 15~fb$^{-1}$ were taken about 40~MeV below
this energy (``off-resonance'') for the study of continuum background in
which a light or charm quark pair is produced instead of an \UfourS.

Charged particles are detected and their momenta measured by the
combination of a silicon vertex tracker, consisting of five layers
of double-sided silicon detectors, and a 40-layer central drift chamber,
both operating in the 1.5-T magnetic field of a superconducting solenoid.
The tracking system covers 92\% of the solid angle in the center-of-mass frame.

Charged-particle identification (PID) is provided by the average
energy loss (\dedx) in the tracking devices and by an internally reflecting ring-imaging
Cherenkov detector (DIRC) covering the central region.
A $K/\pi$ separation of better than four standard deviations ($\sigma$)
is achieved for momenta below 3~\gevc, decreasing to 2.5 $\sigma$ at the 
highest momenta in the $B$ decay final states.

Monte Carlo (MC) simulations of the signal decay modes, 
continuum, \BB\ backgrounds and detector response~\cite{geant4} are used to establish 
the event selection criteria.
The MC signal events are simulated as $B^0$ decays to $a_1(1260) \pi$
with $a_1 \rightarrow \rho \pi$.
For  the  ${a_1(1260)}$ meson parameters we take the mass  
$m_0=1230$ \mevcc\ and  $\Gamma_0=400$ \mevcc\ ~\cite{evtgen,PDG2004} . 

We reconstruct the decay \Natopipipi\ with the
following requirement on the invariant mass: $0.83<m_{\auno}<1.8$~\gevcc.
The intermediate dipion state is reconstructed
with an invariant mass between 0.51 and 1.1~\gevcc. 
We impose  several PID requirements to ensure the
identity of the signal pions. For the bachelor
charged track we require an associated DIRC Cherenkov angle between
$-2\,\sigma$ and $+5\,\sigma$ from the expected value for a pion.
With this requirement all but 1.4\% of any background from
$a_1(1260) K$ is removed.

A $B$ meson candidate is characterized kinematically by the energy-substituted 
mass $\mes = \sqrt{(s/2 + \pvec_0\cdot \pvec_B)^2/E_0^2 - \pvec_B^2}$ and
energy difference $\DE = E_B^*-\half\sqrt{s}$, where the subscripts $0$ and
$B$ refer to the initial \UfourS\ and to the $B$ candidate in the lab-frame, respectively, 
and the asterisk denotes the \UfourS\ frame. The resolutions in \mes\ and 
in \DE\ are  about 3.0  \mevcc\ and  20 \mev\ respectively.
We require $|\DE|\le0.2$ GeV and $5.25\le\mes\le5.29\ \gevcc$. To reduce fake $B$ meson
candidates we require a $B$ vertex $\chi^2$ probability  $>$ 0.01. 
The cosine 
of the angle between the direction of the
$\pi$ meson from \NNatorhopi\   with respect to the flight direction of the $B$ 
in the \auno\ meson rest frame
is required to be between $-0.85$ and $0.85$ to suppress combinatorial
background. The distribution of this variable is flat for signal and peaks near $\pm 1$ for
this background.

To reject continuum background, we use
the angle $\theta_T$ between the thrust axis of the $B$ candidate and
that of the rest of the tracks and neutral clusters in the event, calculated in
the center-of-mass frame. The distribution of $\cos{\theta_T}$ is
sharply peaked near $\pm1$ for combinations drawn from jet-like $q\bar q$
pairs and is nearly uniform for the isotropic $B$ meson decays; we require
$|\cos{\theta_T}|<0.65$. The remaining continuum background is modeled from off-resonance data. 
We use MC simulations of \BzBzb\ and \BpBm\ decays to look for \BB\ backgrounds, 
which can come from both charmless and charm decays. 
We find that the decay mode $B^0 \rightarrow D^- \pi^+$, with 
$D^- \rightarrow K^+ \pi^- \pi^-$ or $D^- \rightarrow K^0_S \pi^-$, 
are the dominant \BB\ backgrounds to ultimate final states different than 
the signal. The decay modes \Nbtoadueppim\ and \Nbtopipppim\  have the same 
final daughters as the signal. We suppress  these with the angular 
variable ${\mathcal A}$, defined as  the cosine
of the angle between the normal to the plane of the $3\pi$ resonance
and the flight direction of the bachelor pion evaluated in the $3\pi$
resonance rest frame. Since the  $a_1(1260)$, $a_2(1320)$ and
$\pi(1300)$ have spins of 1, 2 and 0 respectively,  the distributions of 
the variable ${\mathcal A}$ for these three resonances differ. We require 
$|\mathcal{A}|< 0.62$.

We use an unbinned, multivariate maximum-likelihood fit to extract
the yields of  \Nbtoappim, \Nbtoadueppim\ and \Nbtopipppim.
The likelihood function incorporates five variables.
As mentioned  above, we describe the $B$ decay kinematics with two 
variables: \DE\ and \mes. We also include the invariant mass of the $3\pi$ system, a
Fisher discriminant \xf, and the variable ${\mathcal A}$ (though the
later provides little discrimination after the requirement mentioned above).
The Fisher discriminant combines four
variables: the angles with respect to the beam axis, in the \UfourS\
frame, of the $B$ momentum and $B$ thrust axis, and
the zeroth and second angular moments $L_{0,2}$ of the energy flow
around the $B$ thrust axis. The moments are defined by
\begin{equation}
L_j = \sum_i p_i\times\left|\cos\theta_i\right|^j,
\end{equation}
where $\theta_i$ is the angle with respect to the $B$ thrust axis of
track or neutral cluster $i$, $p_i$ is its momentum, and the sum
excludes tracks and clusters used to build the $B$ candidate.

We have on average 1.4 candidates per event and we select the $B$ 
candidate with the smallest  $\chi^2$ formed from the $\rho$ mass.
The efficiency of the best candidate algorithm is 94\,\%.

Since the correlation between the ob\-ser\-va\-bles in the selected data
and in MC signal events is small, 
% 3.3\,\%, 
we take the probability density function (PDF) for each event to 
be a product of the PDFs for the separate observables.
The product PDF for event $i$ and hypothesis $j$, where
$j$ can be signal \Nappim, \Nadueppim\ and \Npipppim\ backgrounds, 
continuum background or \BB\ background (2 types), is given by:
\begin{equation}
{\cal P}^i_{j} =  {\cal P}_j (\mes) \cdot {\cal  P}_j (\DE) \cdot
{ \cal P}_j(\xf) \cdot {\cal P}_j (m_{a_1}) \cdot {\cal P}_j ({\mathcal A}).
\end{equation}
The probability that inside the signal event the primary pion from the
$B$ candidate is confused with a pion from the $a_1(1260)$ is
negligible because of the high momentum of the primary pion in
\UfourS\ frame.
There is the possibility that a track from a  \Nappim,  \Nadueppim\ and 
 \Npipppim\
event is exchanged with a track from the rest of the event. These 
so-called self cross feed  (SCF) events are considered as background events.
The likelihood function for the event $i$ is defined as
\begin{eqnarray}
{\cal L}^{i} & = & \sum_{k=1}^3 \left(n_k{\cal P}_k + n^{SCF}_k {\cal P}^{SCF}_k\right)
                   + n_{q\bar{q}}{\cal P}_{q\bar{q}}\\
       & + & n_{B\bar{B}1}{\cal P}_{B\bar{B}1} + n_{B\bar{B}2}{\cal P}_{B \bar{B}2}\,, \nonumber
\end{eqnarray} 
where $n_k$ and $n^{SCF}_k$ $(k=1,3)$ are the signal and SCF yields for \Nappim, \Nadueppim,
and \Npipppim, respectively, $n_{q\bar{q}}$ is the number of continuum
background events, $n_{B\bar{B}1}$ is the number of \BB\ background events
$D^-\pi^+$ with $D^- \rightarrow K^+ \pi^- \pi^-$ and $n_{B\bar{B}2}$ is the number 
of \BB\ background events $D^-\pi^+$ with $D^- \rightarrow K^0_S \pi^-$. 
 ${\cal P}_{k}$ is the PDF for correctly reconstructed MC signal events;${\cal P}^{SCF}_k $ is 
the PDF for SCF  events, ${\cal P}_{q \bar{q}}$ is 
the PDF for continuum background events, and ${\cal P}_{B \bar{B}1}$ and ${\cal P}_{B \bar{B}2}$ 
are the PDFs for the two types of $B \bar{B}$ backgrounds, all evaluated with the 
observables of the $i$th event.

We write the extended likelihood function for all events as :
\begin{equation}
{\cal L} = \exp{\Big(-\sum_j n_{j}\Big)}\prod_i^{N}{\cal L}^{i}\,,
\end{equation}
\noindent where $n_{j}$ is the number  of events of hypothesis $j$ found by the
fitter, and $N$ is the number of events in the sample. The first factor takes
into account the Poisson fluctuations in the total number of events.

We determine the PDFs for signal and \BB\ backgrounds from
MC distributions in each observable. For the continuum background we establish
the functional forms and initial parameter values of the PDFs with off-resonance data. 

The PDF of the invariant mass of the ${a_1(1260)}$ meson in  signal events is parameterized as a relativistic Breit-Wigner
line-shape with a mass dependent width which takes into account the effect of the mass-dependent $\rho$ width~\cite{WA76}.
The PDFs of the invariant masses of the  ${a_2(1320)}$ and $\pi(1300)$ mesons are parameterized by triple Gaussian functions.
The \mes\ and \DE\ distributions for 
signal are parameterized as
double Gaussian functions. The \DE\ distribution for continuum background is parameterized by a linear function.
The combinatorial background in \mes\ is described by a phase-space-motivated
empirical function \cite{argus}. We model the
Fisher distribution \xf\ using a Gaussian function with different widths above
and below the mean. The ${\mathcal A}$ distributions are modeled using 
polynomials. 

In the fit there are fourteen free parameters: six yields, the signal ${a_1(1260)}$
mass and width, and six parameters affecting the shape of the combinatorial 
background. Table I lists the results of the final fits. Fitted values
of  ${a_1(1260)}$ mass and width have statistical errors only.

\begin{table}[!htb]
\label{tab:results}
\caption{Signal yield, detection  efficiency ($\epsilon$),  
statistical significance (with systematic 
uncertainties), branching fraction, and the mass and width of the $a_1(1260)$ meson.}
\begin{center}
\begin{tabular}{lc}
\hline
Fit quantity    &\Nappim\  \\
\hline
Signal yield            &$421 \pm 48$ \\
$\epsilon$ (\%) & $11.7 $ \\
Stat. sign. ($\sigma$)   & $9.2$    \\
${\cal B}(\times 10^{-6})$      &$16.6 \pm 1.9 \pm 1.5 $  \\
\hline
$m(a_1(1260))$ & $ 1229 \pm 21$ \mevcc\\
$\Gamma(a_1(1260))$ &$ 393 \pm 62$ \mevcc\\
\hline
\end{tabular}
\end{center}
\end{table}

Equal production rates to \BzBzb\ and \BpBm\ pairs are assumed.
We find no evidence of the decay 
\Nbtopipppim, and therefore we have not included this  component in the final fit. The yield 
of the decay  \Nbtoadueppim\ is $8.3 \pm 23.6$  events.

\begin{figure}[!h]
\resizebox{\columnwidth}{!}{
\includegraphics[]{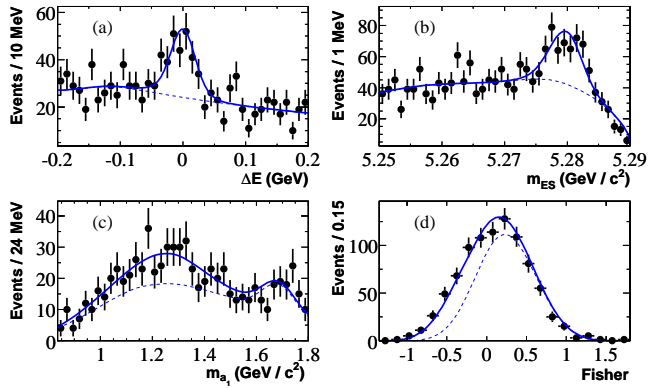} 
}
\caption{Projections of a) \DE, b)  \mes, c) $m_{a_1}$, and d) \xf. 
Points represent on-resonance data, dotted lines 
the continuum and \BB\ backgrounds, and solid lines the full fit
function. These plots are made with a  cut on the signal likelihood
which includes about 40\% of the signal.}
\label{fig:projections}
\end{figure}

We find a signal yield bias of +3.8\,\% by generating and fitting MC
simulated samples containing signal and background populations expected for data.
We find that $-\ln{L}_{\rm{max}}$ from the
on-resonance data lies well within the distribution of $-\ln{L}_{\rm{max}}$ 
from these simulated samples.  
The signal reconstruction  efficiency is obtained from the fraction of signal MC events passing
the selection criteria, adjusted for the bias in the likelihood fit. 
The statistical significance is taken as the square root of the difference between 
the value of $-2\ln{L}$ for zero signal and the value at its minimum.

In Fig.~\ref{fig:projections} we show the \DE, \mes, $m_{a_1}$, and \xf\  
projections made by selecting events with a signal likelihood (computed without the variable
shown in the figure) exceeding a threshold that optimizes the
expected sensitivity. The enhancement at 1.7~\gevcc\ in Fig.~\ref{fig:projections}(c) 
comes from $D^- \pi^+$ background.

\begin{figure}[!htb]
\resizebox{0.8\columnwidth}{!}{
\begin{tabular}{c}
\includegraphics[]{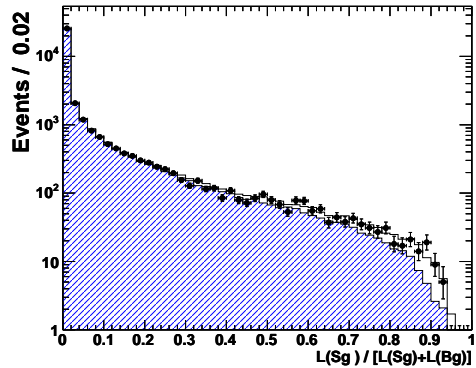}
\end{tabular}
}
\caption{Likelihood ratio L(Sg)/[L(Sg) + L(Bg)]. Points represent 
the data, the solid histogram is from Monte Carlo samples of background  
plus signal, with the background component shaded.  }
\label{fig:LLR}
\end{figure}

In Fig.~\ref{fig:LLR} we show the distribution of the ratio of the 
likelihood for signal events L(Sg) and the sum of likelihoods
for signal and all types of background [L(Sg) + L(Bg)]
for on-resonance data and for Monte Carlo events generated from PDFs.
We see good agreement between the model and the data. By construction
the background is concentrated near zero, while the signal appears as
an excess of events near one. 

Most of the systematic errors on the signal yield that arise 
from uncertainties in the
values of the PDF parameters have already been incorporated into the overall
statistical error, since they are floated in the fit.
We determine the sensitivity to the other parameters of the signal and background PDF
components by varying these within their uncertainties. 

The uncertainty in our knowledge of the efficiency is found
to be 3.2\%.
The systematic error on the fit yield is 6.2\,\%, which is obtained by 
varing the PDF parameters within their uncertainties.
We estimate the uncertainty in the number of \BB\ pairs to be 1.1\,\%.
The uncertainty in the fit bias correction is 1.9\,\%, taken as
half of the fit bias correction. Published world
averages~\cite{PDG2004}\ provide the $B$ daughter branching fraction
uncertainties. The systematic errors on $a_1(1260) K$ cross-feed background
and on SCF are both estimated  to be 1.4\,\%.
The potential background contribution
from $B^0$ decays to $\rho^0 \rho^0$, $\rho^0 \pi^+\pi^-$ and $4\pi$
is estimated assuming the branching fractions of 1, 2, and 2 in
$10^{-6}$ respectively  \cite{RhoRho}. The associated systematic
uncertainty is  3.9\,\%.
The systematic effect due to differences between data and MC for the
\costhr\ selection is 1.8\%. A systematic uncertainty of  2.5\,\% is estimated 
for the difference in reconstruction efficiency in the decay modes through 
the dominant P-wave $(\pi \pi)_{\rho}$ and the S-wave $(\pi
\pi)_{\sigma}$.   The contribution of interference between 
$a_2(1320)$ and $a_1(1260)$ is negligible. In fact, varying 
the $a_2(1320) \pi$ background with different selection criteria on the angular variable 
${\mathcal A}$ gives no significant change to the efficiency-corrected signal yield of $a_1(1260) \pi$.
We find also that the systematic effect due to different form factors in MC signal simulation
is negligible. The total systematic error is 9.1\,\%.

In conclusion, we have measured the branching fraction
\BrapiPiPiPi = \Rapi.
Assuming ${\cal B}(a^{\pm}_1(1260) \rightarrow \pi^{\mp} \pi^{\pm} \pi^{\pm})$ is equal to 
${\cal B}(a^{\pm}_1(1260) \rightarrow \pi^{\pm} \pi^0 \pi^0)$, and that  ${\cal B}(a^{\pm}_1(1260)\rightarrow (3\pi)^{\pm})$ is
equal to 100\%~\cite{PDG2004}, we obtain  ${\cal B}(B^0 \rightarrow
a^{\pm}_1(1260) \pi^{\mp})=(33.2 \pm 3.8 \pm 3.0)\times 10^{-6}$
The decay mode, observed for the first time,  is seen with a significance 
of 9.2~$\sigma$, which includes systematic uncertainties. 

We are grateful for the excellent luminosity and machine conditions
provided by our \pep2\ colleagues, 
and for the substantial dedicated effort from
the computing organizations that support \babar.
The collaborating institutions wish to thank 
SLAC for its support and kind hospitality. 
This work is supported by
DOE
and NSF (USA),
NSERC (Canada),
IHEP (China),
CEA and
CNRS-IN2P3
(France),
BMBF and DFG
(Germany),
INFN (Italy),
FOM (The Netherlands),
NFR (Norway),
MIST (Russia), and
PPARC (United Kingdom). 
Individuals have received support from CONACyT (Mexico), 
Marie Curie EIF (European Union),
the A.~P.~Sloan Foundation, 
the Research Corporation,
and the Alexander von Humboldt Foundation.

\end{document}